\documentclass[prb,showpacs,amsmath,amssymb,twocolumn]{revtex4-1}
\usepackage{graphicx}
\usepackage{amssymb,amsmath}
\usepackage{dcolumn}
\usepackage{bm}
\usepackage{color}

\begin{document}

\title{Bloch-oscillations of exciton-polaritons and photons for the generation of an alternating terahertz spin current}

\author{H. Flayac}
\affiliation{LASMEA, Nanostructure and Nanophotonics Group, Clermont Universit\'{e}}
\affiliation{Universit\'{e} Blaise Pascal, CNRS, 63177 Aubi\`{e}re Cedex France}

\author{D. D. Solnyshkov}
\affiliation{LASMEA, Nanostructure and Nanophotonics Group, Clermont Universit\'{e}}
\affiliation{Universit\'{e} Blaise Pascal, CNRS, 63177 Aubi\`{e}re Cedex France}

\author{G. Malpuech}
\affiliation{LASMEA, Nanostructure and Nanophotonics Group, Clermont Universit\'{e}}
\affiliation{Universit\'{e} Blaise Pascal, CNRS, 63177 Aubi\`{e}re Cedex France}

\begin{abstract}
We analyze theoretically the spin dynamics of exciton-polaritons and photons during their Bloch oscillations in a one-dimensional microwire. The wire geometry induces an energy splitting between the longitudinal and transverse electric polarized eigenmodes. We show analytically that a synchronized regime between the Bloch oscillations in space and the precession of the pseudospin can be achieved. This synchronization results in the formation of a THz alternating spin current which can be extracted out of the confinement region by the Landau-Zener tunneling towards the second allowed miniband. Finally we show how the spin signal can be maintained despite of the lifetime of the particles for the case of exciton-polaritons both in a resonant and non resonant pumping schemes. The structure therefore acts as a Spin-optronic device able to convert the polarization and emit spin polarized pulses.

\end{abstract}

\maketitle

\section{Introduction}
Exciton-polaritons (polaritons)\cite{Microcavities} are now attracting a lot of attention since they have demonstrated numerous fascinating nonlinear and linear physical effects. Polaritons are the quasi-particles resulting from the strong coupling between excitons confined in quantum wells and photons confined in a planar Fabry-Perot resonator. Because of their photonic part, polaritons are efficiently emitting and absorbing light. Polaritons are also very light quasi-particles which can very propagate inside the cavity with large in-plane velocities. Because of their excitonic part, polaritons are not only interacting with each other but also with the crystal lattice (phonons). The self interaction property gives the polaritonic system an extremely high non-linear optical response. This was first evidenced by the demonstration of polariton parametric amplification\cite{Savvidis, Stevenson} and bistability\cite{Baas, Gippiusb} of the polariton system. The threshold of these processes is orders of magnitude smaller than the one required to achieve similar effects in standard non-linear optical media\cite{Baumberg}. Besides, the polariton-phonon interactions allow to thermalize the polariton gas, which has led to the achievement of bosonic phase transitions such as Bose Einstein condensation \cite{BECPolaritons}. The latter can be used as a base to implement a very low threshold coherent light emitter, the so-called polariton laser\cite{PolaritonLaser}. The opportunity offered by the system itself and also by the very powerful techniques of optics available nowadays has allow to demonstrate a wide range of quantum fluids phenomena, such as superfluid-like behavior\cite{AmoNatPhys}, quantized vortices\cite{LagoudakisV}, and oblique solitons\cite{AmoSoliton} formation.

Another remarkable property of polaritons is the presence of two spin components. A large variety of original spin-dependent phenomena have been predicted and observed. One can mention the observation of the optical spin-Hall effect\cite{OSHEth,OSHEexp}, the polariton multistability \cite{Gippiusm, Gippiusm2} or half vortices \cite{Rubo,LagoudakisHV}. The polariton physics therefore clearly lies at the crossroads of several important fields of physics which are non-linear optics, Bose-Einstein condensation and spintronics. Fundamental physics merges here with the applied one. The most well known polariton devices are polariton lasers and amplifiers which do not require the achievement of the gain condition needed in standard "photon" devices. The pumping threshold required for polariton devices is therefore typically at least one order of magnitude lower \cite{LasingThreshold,Ferrier2011}. Many proposals of new types of optical spin-based microscopic components (spin-optronic devices) have been made in the past years \cite{review_Shelykh}. Non-exhaustively, one can cite terahertz sources\cite{Terahertz}, optical circuits based on neurons\cite{Neurons}, optical gates\cite{OpticalGate}, Berry phase interferometer\cite{BerryPhase}, spin transistor\cite{DattaDas}, or spin switches \cite{Lausanne}.

Recently we have predicted the possibility for polaritons to achieve a fundamental quantum phenomenon: Bloch oscillations\cite{FlayacBO} (BOs) within a wire-shaped microcavity. BOs rely on the action of a constant force (varying lateral size of the wire or cavity thickness) on the particles in the presence of a periodic potential (metallic pattern depositions, surface acoustic waves, square-wave lateral etching...). Due to Bragg reflections at the first Brillouin zone (FBZ) edges, particles oscillate instead of infinitely accelerating. This effect has been described in many other systems and has been proposed as a possible source of terahertz radiation\cite{TerahertzBOs}. BOs could also be used to measure very sensitive atomic physical quantities \cite{CarusottoSForce,Gravity}. In this paper we show that BOs of polaritons can be used in their case to implement an ultra-fast spin switch/converter or even a spin transistor. We will first mostly concentrate on the low-density linear regime, neglecting polariton-polariton interactions. In such a regime, low-momentum polaritons do not strongly differ from cavity photons, and most of the effects we find can perfectly be obtained for a purely photonic system\cite{MalpuechBO,AgarwalBO} as it was recently the case for the optical spin Hall effect \cite{Plagoud}. In the very last part we will show how the polariton signal can be maintained (stimulated) thanks to their nonlinear dynamics and their interaction with a reservoir produced by a non-resonant pumping.

We propose to consider a patterned microwire similar to the one described in Ref.\onlinecite{FlayacBO}. Such a structure imposes a specific longitudinal/transverse (LT) splitting useful to manipulate the polariton's (photon's) polarization. We first demonstrate analytically that polarization precession and BOs period can be synchronized. One half of the spatial oscillations is performed by one spin component and the second half by the other. This regime could therefore be somehow called "half Bloch-oscillations". In the second part of the paper, we demonstrate that this synchronized regime combined with the Landau-Zener tunneling (LZT) to the second Bloch band allows THz transmission of picosecond pulses with alternating circular polarization. No external magnetic field needs to be applied to the cavity. In such case, the polariton and photon behavior in the low wave vector region are very similar. The disadvantage of polaritons is mostly that they of course require low temperature operation. Their advantage in the linear regime lies in the possibility to finely tune the energy of the polarized polariton modes\cite{MalpuechAPL} with an electric field acting on the quantum well excitons. Similar modulation can of course be achieved on purely photonic systems with the Pockels or Kerr effects, but they require field intensities orders of magnitude larger than in the polariton systems. We will describe the polaritonic system as a reference and, when needed, comment on analogies with photonic systems.

\section{Spin structure and LT splitting}
Polaritons are bosons that have an electron-like two-level spin structure\cite{review_Shelykh}. Indeed, only specific excitons that have $\sigma_{\pm}=\pm1$ spin projection are able to couple to cavity photons (bright states), in contrast to the dark states (spin $\pm2$). A right (left) circularly polarized light excitation creates a $\sigma_+$ ($\sigma_-$) polariton. Consequently, a convenient representation for the spin (polarization) dynamics of the polaritonic system is the three dimensional pseudospin vector (analogue of the Stokes vector) $\mathbf{S}=(S_x,S_y,S_z)$ on the Poincar\'{e} sphere. This vector is defined as the the decomposition of the $2\times2$ spin-density matrix $\rho_s$ of polaritons on a set consisting of the unity matrix $\mathbf{I}$ and the three Pauli matrices $\sigma_{x,y,z}$: $\rho_s= \mathbf{I}N/2 + \mathbf{S} \cdot \boldsymbol{\sigma}$ with $N$ the total number of particles in the system. The pseudospin allows us to map the system to a magnetic one and completely defines the polarization of the system: the $S_x$ and $S_y$ describe the linear polarization states while the $S_z$ component is the circular polarization degree of the particles when normalized to unity.

In microcavities, optical eigenmodes are Transverse Electric (TE) and Transverse Magnetic (TM). For a two dimensional system, the energy splitting between these modes grows quadratically with the wave vector $\mathbf{k}$. The resulting polariton states are also TE and TM polarized. Another important contribution to the polariton TE-TM splitting is the opposite dependence of the exciton coupling strength with the TE and TM modes. This contribution, which is usually negligible in the small $\mathbf{k}$ range where the strong coupling is taking place, is due to the long range interaction between the electrons and the holes. For excitons having nonzero in-plane wave vectors, the eigenstates with dipole moment oriented along (TM) and perpendicular (TE) to $\mathbf{k}$ are slightly different in energy\cite{Maialle} for any $\mathbf{k}\neq0$. The TE-TM splitting is an important feature: It acts on polaritons pseudospin and manifests as an effective magnetic field that lies in the plane of the cavity and makes a double angle with the propagation direction of the particles
\begin{equation}\label{OmegaLT1}
    {\boldsymbol{\Omega}_{LT}}\left(\mathbf{k}\right) = \omega_x\left(\mathbf{k}\right)\cos \left( {2\phi } \right){\mathbf{u}_{x}} + \omega_y\left(\mathbf{k}\right)\sin \left( {2\phi } \right){\mathbf{u}_{y}}
\end{equation}
$\phi$ is the polar angle. These peculiarities result in the precession of the pseudospin providing a remarkable spin dynamics and related phenomena such as the OSHE, the formation of polarization patterns\cite{PolPatterns} or oblique half-solitons \cite{FlayacHS}.

\begin{figure}
  \includegraphics[width=0.45\textwidth,clip]{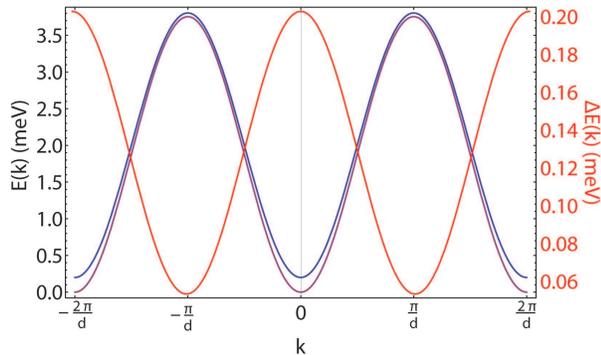}\\
  \caption{(Color online) First TE (solid blue line and left scale) and TM (solid purple line and left scale) Bloch bands and their energy splitting (solid red line and right scale) for the parameters defined in Sec.III.}
  \label{Fig00}
\end{figure}

Moreover, in quasi-one dimensional microcavities the TE and TM eigenmodes are linearly polarized perpendicular and parallel to the wire's axis ($x$-axis) respectively. The additional confinement lifts the degeneracy between the TE and TM modes even at $\mathbf{k}=\mathbf{0}$ like in usual photonic waveguides. It induces an additional effective magnetic field along the $x$-axis. This splitting is already present in planar structures as first demonstrated in \cite{Klop}, it is however much larger in wires. Mainly because of strain relaxation, the effective values can moreover be much larger than the one extracted from Maxwell's equations in isotropic media\cite{Dasbach}. An advantage of the polaritonic system over the purely photonic one is that the energy of the exciton state coupled to one polarization or another can be finely tuned by applying a moderate electric field \cite{MalpuechAPL}. This can be used to achieve the synchronization between BOs and the polarization rotation, as we will show below. In what follows, we will consider wires similar to the one studied in Ref.\onlinecite{Wertz}. In these samples, the total energy splitting is the strongest at $\mathbf{k}=\mathbf{0}$ and diminishes for increasing $\mathbf{k}$.
As described in Ref.\onlinecite{FlayacBO}, the addition of a periodic pattern leads to a band-structured dispersion of the polaritons. The first TE and TM bands as well as their energy splitting gain a $2\pi/d$ periodicity, where $d$ is the period of the patterned potential. In what follows, we will see how it influences the spin dynamics (pseudospin precession) of the system. We show in Fig.\ref{Fig00} the corresponding first TE and TM bands and the energy splitting between the two for a GaAs microwire, assuming a strong periodic potential for simplicity. The parameters are those given in the Section \ref{sec3}.

\section{Spin dynamics induced by the polarization splitting}\label{sec3}
\begin{figure}
  \includegraphics[width=0.4\textwidth,clip]{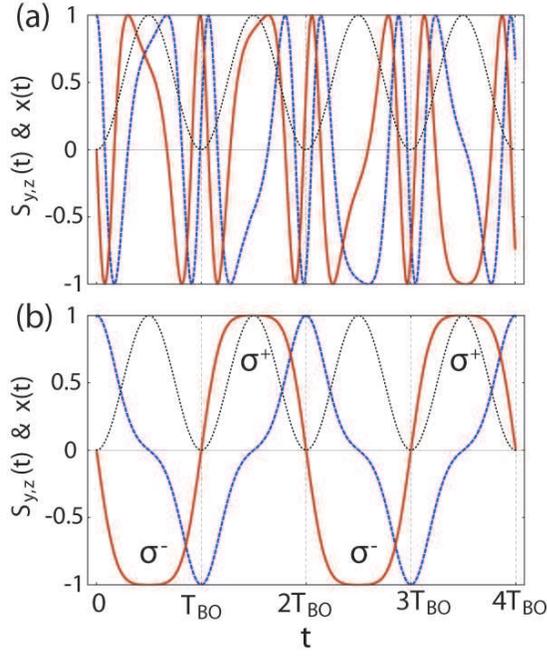}\\
  \caption{(Color online) Pseudospin dynamics during BOs (a) for arbitrary parameters and (b) following the condition (\ref{Criterion}). The solid red curves shows the normalized circular polarization degree ($S_z$), the dashed blue curves shows the $S_y$ component ($S_x$ is always zero) and the dotted black line stands for the trajectory of the center of mass of the wave packet.}
  \label{Fig0}
\end{figure}

\begin{figure}
  \includegraphics[width=0.5\textwidth,clip]{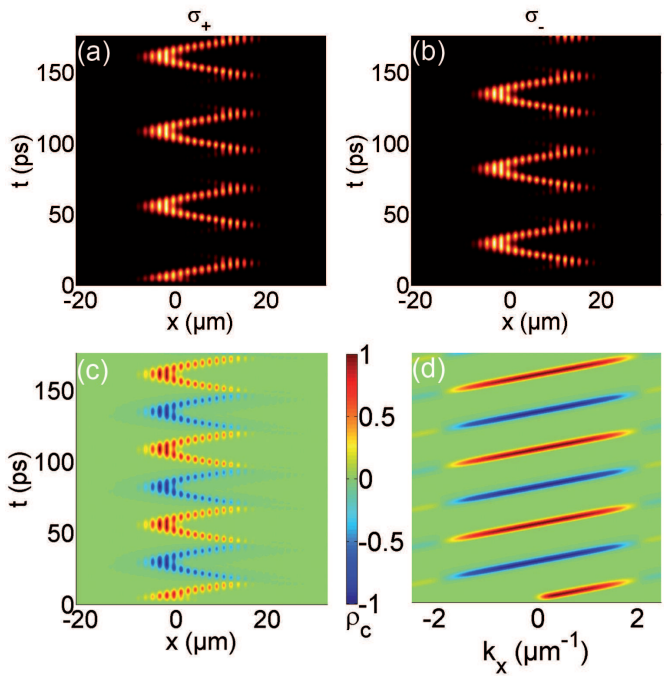}\\
  \caption{(Color online) Half Bloch-Oscillations. The (a) and (b) panels show the emission intensity in real $\sigma_+$ and $\sigma_-$ space respectively. (c) and (d) are the corresponding circular polarization degree in real and momentum space respectively. Parameters are given in the text.}
  \label{Fig1}
\end{figure}

In the linear regime, the dynamics of the pseudospin of the center of mass of a wave packet in the presence of an effective magnetic field associated with the energies $\hbar \boldsymbol{\Omega}$ and neglecting any dissipation is given by the following vectorial equation
\begin{equation}\label{Seq}
    {\partial _t}\mathbf{S}\left(t \right)  = \mathbf{S}\left(t \right)  \times {\boldsymbol{\Omega}}
\end{equation}
In the first part of this section we assume for simplicity that our system is described by a tight binding approximation. We therefore consider cosine shaped bands and a strictly one dimensional system ($k_x\rightarrow k$). This approximation is reasonable insofar we will discuss a phenomenon linked with the period of the BOs (which doesn't depend on the width of the band) and not with their amplitude. Under such conditions, the first Bloch bands (see Fig.\ref{Fig00}) for the TE and TM states are given by $E_l(k)=J_{l}[1-\cos(k d)]$ and $E_t(k)=J_{t}[1-\cos(k d)]+H_x$, where $H_x$ accounts for the energy splitting at $k=0$ (static in plane field) and $J_{t,l}$ are the coupling constants between adjacent wells. They can be approximated\cite{ReviewAtoms} by $J_{t,l}=4\varepsilon_{t,l}(A/\varepsilon_{t,l})^{3/4}\exp(-2\sqrt {A/\varepsilon_{t,l}})$ with $\varepsilon_{t,l}={\hbar^2}{\pi^2}/2{m_{t,l}}{d^2}$ the recoil energies and $A$ the amplitude of the periodic potential. Then, the $k$-dependence of the effective field along the wire reads
\begin{equation}\label{OmegaLT}
    {{\boldsymbol{\Omega }}_{LT}}=\Omega_x \mathbf{u}_{x}= \frac{{{H_x}-\Delta J\left[ {1 - \cos \left( {kd} \right)} \right]}}{\hbar } \mathbf{u}_{x}
\end{equation}
with $\Delta J=J_l-J_t$. Eq.\ref{Seq} leads to the following coupled equations for the evolution of the pseudospin components:
\begin{eqnarray}
{\partial _t}{S_x}\left( t \right) &=& 0\\
{\partial _t}{S_y}\left( t \right) &=&  + {\frac{{{H_x}-\Delta J\left[ {1 - \cos \left( {k\left( t \right)d} \right)} \right]}}{\hbar }}{{{S}}_z}\left( t \right)\\
{\partial _t}{S_z}\left( t \right) &=&  - {\frac{{{H_x}-\Delta J\left[ {1 - \cos \left( {k\left( t \right)d} \right)} \right]}}{\hbar }}{{{S}}_y}\left( t \right)
\end{eqnarray}
Under the action of a constant force $F$, the particles (an input gaussian pulse) exhibit Bloch oscillations. Therefore, in the equations for the pseudospin, $k$ has, of course, to be time dependent to take into account the motion of the center of mass of the wave packet $k(t)=F t/\hbar$, and the period of oscillations depends on the splitting $F d$ between Wannier-Stark states: $T_{BO}=2 \pi \hbar/F d$. Putting for example Eq.(6) into Eq.(5) gives a decoupled equation for $S_z$ and then $S_y(t)$ is completely defined by the knowledge of $S_z(t)$. Finally, we obtain the following pseudospin dynamics
\begin{widetext}
\begin{eqnarray}
{S_x}\left( t \right) &=&  + {S_{0x}}\\
{S_y}\left( t \right) &=&  + {S_{0y}}\cos \left[ {\frac{{\left( {{H_x} - \Delta J} \right)}}{\hbar }t + \frac{{\Delta J}}{\hbar }\frac{{{T_{BO}}}}{{2\pi }}\sin \left( {\frac{{2\pi }}{{{T_{BO}}}}t} \right)} \right] + {S_{0z}}\sin \left[ {\frac{{\left( {{H_x} - \Delta J} \right)}}{\hbar }t + \frac{{\Delta J}}{\hbar }\frac{{{T_{BO}}}}{{2\pi }}\sin \left( {\frac{{2\pi }}{{{T_{BO}}}}t} \right)} \right]\\
{S_z}\left( t \right) &=&  - {S_{0y}}\sin \left[ {\frac{{\left( {{H_x} - \Delta J} \right)}}{\hbar }t + \frac{{\Delta J}}{\hbar }\frac{{{T_{BO}}}}{{2\pi }}\sin \left( {\frac{{2\pi }}{{{T_{BO}}}}t} \right)} \right] + {S_{0z}}\cos \left[ {\frac{{\left( {{H_x} - \Delta J} \right)}}{\hbar }t + \frac{{\Delta J}}{\hbar }\frac{{{T_{BO}}}}{{2\pi }}\sin \left( {\frac{{2\pi }}{{{T_{BO}}}}t} \right)} \right]
\end{eqnarray}
\end{widetext}
where $S_{0i}=S_{i}(0)$. This solution is deterministic with respect to the sample parameters: a given initial pseudospin vector $\mathbf{S}_0=(S_{0x},S_{0y},S_{0z})$ (the polarization of the input pulse) completely defines the spin dynamics of the system. As we are dealing with the evolution of a single particle (the center of mass of a wave packet), the pseudospin vector $\mathbf{S}$ should be normalized to unity, it imposes: $S_{0x}^2+S_{0y}^2+S_{0z}^2=1$. The maximum precession amplitude given by Eq.\ref{Seq} is obtained for $\mathbf{S}\perp \boldsymbol{\Omega}_{LT}$. For arbitrary parameters, the precession of $\mathbf{S}$ is expected to be unsynchronized with $T_{BO}$ as we can see in the Figure \ref{Fig0}(a). Now, we can impose a specific pseudospin state $\mathbf{S}_{BO}=\mathbf{S}(j T_{BO})$ with $j$ an integer (actually $j=1$ is sufficient), in order to synchronize the pseudospin precession with the oscillations of the wave packet. We show in Fig.\ref{Fig1}(b) an example of such a synchronization regime with the set of conditions $\mathbf{S}_0=(0,+1,0)$ (diagonal linear polarization) and $\mathbf{S}_{BO}=(0,-1,0)$ (anti-diagonal). The corresponding synchronization criterion reads
\begin{equation}\label{Criterion}
    {T_{BO}}{\rm{ }} = \frac{{\pi \hbar \left( {1 + 2\kappa } \right)}}{{{H_x} - \Delta J}} \Leftrightarrow F = \frac{{2\left( {{H_x} - \Delta J} \right)}}{{\left( {1 + 2\kappa } \right)d}}
\end{equation}
where $\kappa$ is an integer taken to be zero for the case of Fig.\ref{Fig0}(b). For example, using the typical parameters $H_x=0.2$ meV and $\Delta J=0.1$ meV we obtain $T_{BO}\simeq20$ ps and $F\simeq0.13$ meV$/\mu$m which enters perfectly in the range of accessible values for polaritonic or photonic systems.

Let us now switch to a numerical modeling of the system. For this purpose, we use a set of spin dependent Schr\"{o}dinger equations and for a first simple description, we start by neglecting the lifetime of the particles and assume parabolic bare dispersions.
\begin{widetext}
\begin{equation}
i\hbar \frac{\partial }{{\partial t}}\left( \begin{array}{l}
{\psi _ + }\\
{\psi _ - }
\end{array} \right) = \left( {\begin{array}{*{20}{c}}
{ - \frac{{{\hbar ^2}}}{{2{m^*}}}\frac{{{\partial ^2}}}{{\partial {x^2}}} + U}&{\beta \frac{{{\partial ^2}}}{{\partial {x^2}}} + {H_x}}\\
{\beta \frac{{{\partial ^2}}}{{\partial {x^2}}} + {H_x}}&{ - \frac{{{\hbar ^2}}}{{2{m^*}}}\frac{{{\partial ^2}}}{{\partial {x^2}}} + U}
\end{array}} \right)\left( \begin{array}{l}
{\psi _ + }\\
{\psi _ - }
\end{array} \right) + \left( \begin{array}{l}
{P_ + }\\
{P_ - }
\end{array} \right)
\end{equation}
\end{widetext}
This first description suits well a pure photonic system and is the common simplest approximation to the polaritonic system which will be extended in Sec.\ref{sec5}. The initial Gaussian light pulse injected via $P_{\pm}(x)$ is right circularly polarized ($\mathbf{S}=(0,0,+1)$), resonant with the lower polariton branch (LPB) at $k=0$ and its amplitude is taken low enough to consider a linear regime. The effective mass is defined by $m^*=2 m_t m_l/(m_t+m_l)$ where $m_t=5\times10^{-5} m_0$, $m_l=0.95 m_t$ are the masses of the transverse and longitudinal modes and $m_0$ is the free electron mass. We note that the mass of the polariton is usually of the order of twice the cavity photon mass. $U(x)$ is the total external potential: The sum of the squarewave periodic potential of amplitude $A=5$ meV (large enough to stay close to the tight binding approximation) and period $d=1.56$ $\mu$m and a ramp potential $-F x$,  $F=0.1$ meV/$\mu$m$^{-1}$ being a constant force. The off diagonal terms accounts for the $k$-dependent LT splitting, where $\beta=\hbar^2/4(m_l-m_t)/(m_l m_t)$. We remind that the components of $\mathbf{S}$ can be explicitly defined via the wave functions $\psi_\pm$.
\begin{eqnarray}
\nonumber {S_x} &=& {\mathop{\Re}\nolimits} \left( \psi _{+} \psi_-^*\right)\\
{S_y} &=& {\mathop{\Im}\nolimits} \left( \psi _{-} \psi_+^*\right)\\
\nonumber {S_z} &=& \left({{\left| {{\psi _ + }} \right|^2} } - {{\left| {{\psi _ - }} \right|^2}}\right)/2
\end{eqnarray}
The Figure \ref{Fig1} shows the probability density of the $\sigma^+$ and $\sigma^-$ components in real space in (a) and (b) respectively. In (c)-(d) we plot the degree of circular polarization, which is nothing but $S_z$ because it is normalized to unity, in real and momentum space respectively. Remarkably, as described analytically in the previous section, every single spatial oscillation in the first Brillouin zone displays alternatively a right or left circular polarization. Because each spin component is present in the system only for a half-period, we will call this regime "Half-Bloch Oscillations" (HBOs).

\section{Emission of a spin signal}
\begin{figure}
  \includegraphics[width=0.5\textwidth,clip]{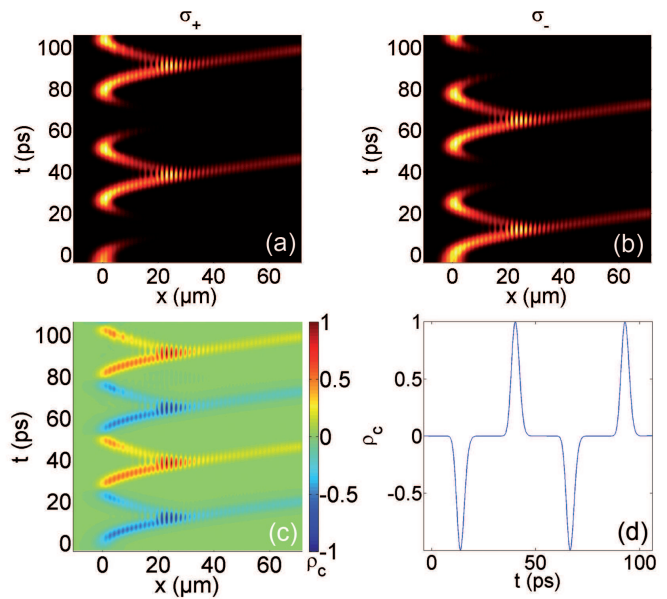}\\
  \caption{(Color online) LZT emission regime. (a), (b) and (c) show the same informations as Fig.\ref{Fig1} (in (a) and (b) the local density is normalized for clarity) while the (d) panel shows the normalized circular polarization degree of the emitted signal $40$ $\mu$m away from the input pulse. The latter has been filtered so that very low density region don't contribute to the signal.}
  \label{Fig2}
\end{figure}

So far, we have been working with wide gaps between the minibands which is not completely realistic for the case of polaritons because of the limitations on the height of the periodic potential imposed by technological constraints\cite{FlayacBO}. When a particle lying in the lowest Bloch band is accelerated up to the FBZ borders, there is a finite probability $P_{LZT} \simeq \exp \left( { - {A^2}m^*\pi /2F} \right)$ of the transition towards the second miniband. This effect is known as the Landau-Zener tunneling (LZT) and induces a signal loss every single oscillation. This is usually harmful for the observation of steady-state BOs. We are going to take advantage of this effect in order to generate periodic polarized light beams at the FBZ edges. Indeed, reducing  the value of the periodic potential's amplitude $A$ to a more realistic value will tend to increase $P_{LZT}$ and then induce a significant emission at every oscillations where the band separation is the smallest (at the FBZ edges). The peculiarity of our spin dependent system is that the emitted pulses will have a specific circular polarization degree controlled by the coupling between BOs and the pseudospin precession. Indeed the LZT occurs every $j^{th}+1/2$ oscillations and the corresponding emission has a circular polarization degree
\begin{eqnarray}
\nonumber S_z^{LZT} =  &+& {S_{0y}}\sin \left[ {\left( {j + 1} \right)\frac{{{T_{BO}}}}{2}\frac{{\left( {{H_x} - \Delta J} \right)}}{\hbar }} \right]\\
 &-& {S_{0z}}\cos \left[ {\left( {j + 1} \right)\frac{{{T_{BO}}}}{2}\frac{{\left( {{H_x} - \Delta J} \right)}}{\hbar }} \right]
\end{eqnarray}
In particular, using the same conditions as in Fig.\ref{Fig0}(b), the normalized circular polarization degree of the emitted signal is ${S_z^{LZT}}=(-1)^{j+\kappa}$. We show in Fig.\ref{Fig2} a synchronized configuration for $A=1$ meV, $\kappa=0$ and $H_x=0.2$ meV. The LZT induced signal measured $40$ $\mu$m away from the input pulse reveals an alternating spin current between $\sigma_+$ and $\sigma_-$. We make the following remark: of course, if the effective magnetic field is present along the whole wire, the signal's pseudospin is expected to keep on rotating while it propagates, which can either be regarded as an issue or not. In such case the polarization of the output signal will crucially depend on the propagation distance in the sample. However, in the synchronized regime, the relative polarization between two consecutive pulses will not depend on the distance, therefore it can be regarded as the real quantity to be measured. Since the effective field depends strongly on the wire lateral size, it can be reduced in the region of free propagation, so that it will not affect the polarization of the emitted signal significantly during its propagation time. In the Figure \ref{Fig2}, $\boldsymbol{\Omega}_{LT}$ is acting only in the BOs region in order to preserve the polarization of the signal for the sake of clarity.

\section{Realistic polaritonic system}\label{sec5}
In this section we will focus on the polariton system. To accurately describe the particle's dynamic, we include in our model both the real dispersion of the particles and their lifetime. We thus introduce the following set of four coupled equations for the excitonic $\psi_\pm^{ph}$ and photonic $\psi_\pm^{ex}$ fields coupled by the light-matter interaction
\begin{widetext}
\begin{eqnarray}
i\hbar \frac{{\partial \psi _ \pm ^{ph}}}{{\partial t}} &=&  - \frac{{{\hbar ^2}}}{{2{m_{ph}}}}\frac{{{\partial ^2}\psi _ \pm ^{ph}}}{{\partial {x^2}}} + \frac{{{\Omega _R}}}{2}\psi _ \pm ^{ex} - \frac{{i\hbar }}{{2{\tau _{ph}}}}\psi _ \pm ^{ph} + {U_{ph}}\psi _ \pm ^{ph} + \left( {\beta \frac{{{\partial ^2}}}{{\partial {x^2}}} + {H_x}} \right)\psi _ \mp ^{ph} + {P_ \pm }\\
i\hbar \frac{{\partial \psi _ \pm ^{ex}}}{{\partial t}} &=&  - \frac{{{\hbar ^2}}}{{2{m_{ex}}}}\frac{{{\partial ^2}\psi _ \pm ^{ex}}}{{\partial {x^2}}} + \frac{{{\Omega _R}}}{2}\psi _ \pm ^{ph} - \frac{{i\hbar }}{{2{\tau _{ex}}}}\psi _ \pm ^{ex} + {U_{ex}}\psi _ \pm ^{ex}
\end{eqnarray}
\end{widetext}
The new quantities that appear are: the effective masses for the cavity photons $m_{ph}=5\times10^{-5}m_0$ and the excitons $m_{ex}=0.5 m_0$, their lifetimes $\tau_{ph}=50$ ps and $\tau_{ex}=400$ ps, the separated\cite{FlayacBO} periodic potential $U_{ex}$ of amplitude $A=1$ meV (acting on excitons) and the ramp potential $U_{ph}$ (acting on photons) which produces a force $F=0.2$ meV$/\mu$m and the Rabi splitting $\Omega_{R}=14$ meV. The analytical description of the previous section appears of course a bit less accurate with respect to the full treatment. The dependence of $\boldsymbol{\Omega}_{LT}$ over $k$ will be slightly affected because of the modified shape of the first Bloch band which mostly modifies the amplitude of oscillations and the shape of the pseudospin oscillations but is not detrimental for our effect. The period $T_{BO}$ is not expected to vary to much because it depends on the quantity $F d$, and therefore, our synchronization criterion remains valid. The design of a real sample would require of course a comprehensive description of the dispersion imposed by the structure which has to be predicted by full two dimensional simulations.

\begin{figure}
  \includegraphics[width=0.5\textwidth,clip]{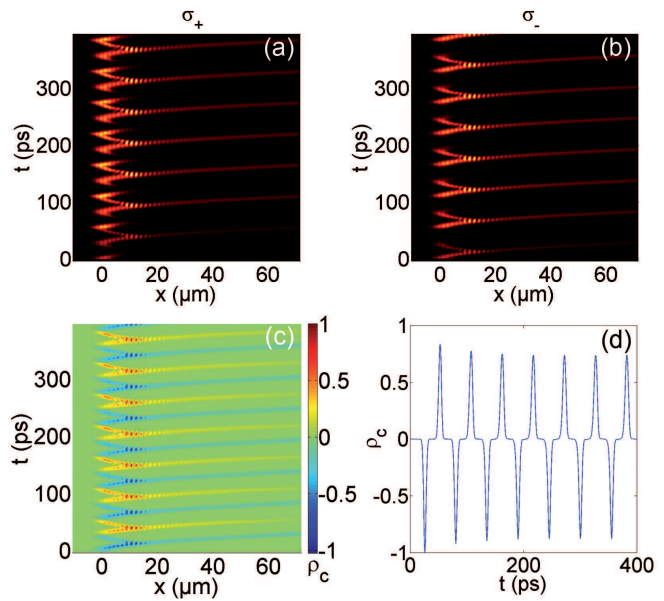}\\
  \caption{(Color online) Accounting for the real non-parabolic dispersion and the lifetime of the particles. Input pulses with a period of $2 T_{BO}\simeq50$ ps are used to maintain the output signal intensity.}
  \label{Fig3}
\end{figure}

We aim now at reproducing the synchronized LZT emitter regime of Fig.\ref{Fig2}. To do so, we need to compensate for the particles losses. It can easily be done thanks to a pulsed input synchronized with $T_{BO}$ as seen in Fig.\ref{Fig3} where the effect is reproduced on demand every two oscillations periods ($T_{BO}=250$ ps) when the signal becomes weakens too much due to both LZT and lifetime. The wire is therefore acting as an ultrafast spin emitter which converts a linearly polarized input in two oppositely circularly polarized outputs in that particular case. Many other configurations are also possible, depending on the parameters imposed by the sample and the polarization of the input, for example the conversion from circular to linear polarization.

Finally, let us consider a sample etched specifically to achieve the synchronization regime described above. A small controllable perturbation to the BOs period or to the LT splitting, produced by an electric contact \cite{MalpuechAPL} or even strain on the sample, will lead to a loss of this synchronization and to an arbitrary relative circular polarization degree between two consecutive LZT pulses. To sum up in the synchronization regime the output is $\Delta{S_z}=2$ and can stand for a binary $\mathbf{1}$ and any unsynchronized configuration is $\mathbf{0}$ with a switching controlled by a gate (perturbations). With this we have described a spin-optronic transistor working at the frequency $1/T_{BO}$ in the range of tens of terahertz.

\section{Driven Half-Bloch oscillations and LZT emission}
In this last section we propose a scheme that will allow us to maintain the polaritons Bloch oscillations and the spin signal discussed previously with a single triggering pulse despite of the lifetime of the particles. This section can also be seen as an extension of Ref.\onlinecite{FlayacBO} where we considered free decaying oscillations of polaritons (condensate).

So far we have been discussing quasi-resonant injections of polaritons (photons) in a linear regime. We propose now to continuously and non-resonantly pump\cite{Wertz} the system with a narrow and high power Gaussian spot in the region of the LZT emission in order to introduce a driving reservoir for the particles. We trigger Bloch oscillations with a single quasi-resonant pulse. To model the action of the reservoir, we need an extra equation for its dynamic and to take particles interaction into account. For the full nonlinear dynamics of the system we therefore use the following set of modified Ginzburg-Landau equations similarly to what was done in Ref.\onlinecite{Keel} for example.

\begin{widetext}
\begin{eqnarray}
i\hbar \frac{{\partial \psi _ \pm ^{ph}}}{{\partial t}} &=&  - \frac{{{\hbar ^2}}}{{2{m_{ph}}}}\frac{{{\partial ^2}\psi _ \pm ^{ph}}}{{\partial {x^2}}} + \frac{{{\Omega _R}}}{2}\psi _ \pm ^{ex} - \frac{{i\hbar }}{{2{\tau _{ph}}}}\psi _ \pm ^{ph} + {U_{ph}}\psi _ \pm ^{ph} + \left( {\beta \frac{{{\partial ^2}}}{{\partial {x^2}}} + {H_x}} \right)\psi _ \mp ^{ph} + {P_ \pm }\\
i\hbar \frac{{\partial \psi _ \pm ^{ex}}}{{\partial t}} &=&  - \frac{{{\hbar ^2}}}{{2{m_{ex}}}}\frac{{{\partial ^2}\psi _ \pm ^{ex}}}{{\partial {x^2}}} + \frac{{{\Omega _R}}}{2}\psi _ \pm ^{ph} - \frac{{i\hbar }}{{2{\tau _{ex}}}}\psi _ \pm ^{ex} + {U_{ex}}\psi _ \pm ^{ex} + \alpha \left( {{{\left| {\psi _ \pm ^{ex}} \right|}^2} + {n_R}} \right)\psi _ \pm ^{ex} + \frac{{i{\Gamma _R}}}{2}{n_R}\psi _ \pm ^{ex}\\
\frac{{\partial {n_R}}}{{\partial t}} &=& {P_R} - \frac{{{n_R}}}{{{\tau _R}}} - {\Gamma _R}\left( {{{\left| {\psi _ + ^{ex}} \right|}^2} + {{\left| {\psi _ - ^{ex}} \right|}^2}} \right){n_R}
\end{eqnarray}
\end{widetext}

\begin{figure}
  \includegraphics[width=0.5\textwidth,clip]{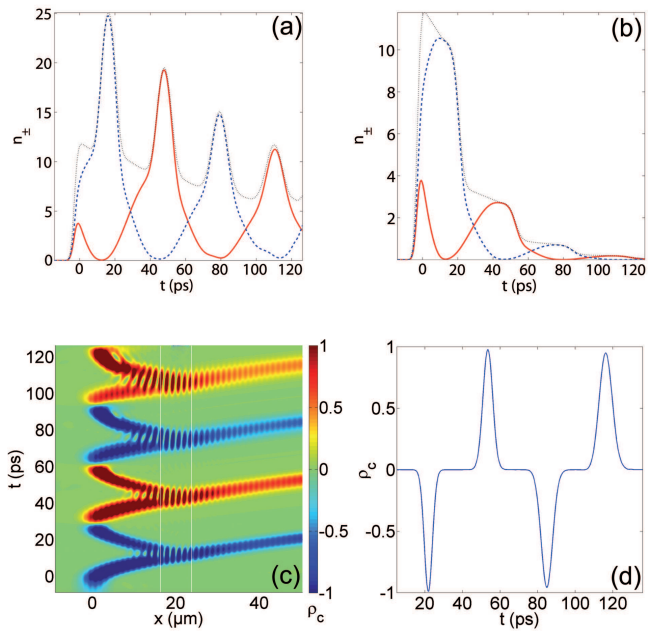}\\
  \caption{(Color online) Half-Bloch oscillations stimulated by a local reservoir. Density of photonic particles for driven Bloch oscillations (a) compared to free oscillations (b) with a 50 ps lifetime. The solid-red and dashed-blue curve stand for the total density of $\sigma_+$ and $\sigma_-$ particles respectively versus time and the dotted black curve shows the sum of the two. (c) Degree of circular polarization in real space, the solid white lines show the position of the reservoir. (d) LZT spin signal emitted.}
  \label{Fig6}
\end{figure}

This model has the advantage of being reasonably simple but does not take into account the interaction with surrounding phonons excitations and therefore any type of thermalization. These considerations could be treated via a master equation approach\cite{MagnussonJO,SavenkoDM} but are not in the focus of the present paper. We assume an exciton reservoir with a lifetime $\tau_R=500$ ps which population $n_R$ evolves along Eq.(18) and populated by a non-resonant localized \emph{cw}-pump $P_R(x)=A_R\exp[(x-x_R)^2/\sigma_R^2]/\tau_R$ with $x_R=25$ $\mu$m and $\sigma_R=2$ $\mu$m. $\Gamma_R=200/\tau_{R}$ the scattering rate towards the polariton condensate. The interactions between particles with parallel spins are introduced via the constant $\alpha=6 E_{b} a_{B}^2/S$, where $E_b=10$ meV is the exciton binding energy, $a_B=10^{-2}$ $\mu$m its Bohr radius, and $S$ is the normalization area. For the pumping value we consider, the presence of the reservoir induces an effective potential barrier $\alpha n_R(x)\sim1$ meV. This moderate value perturbs only weakly the oscillations of the wave packet. We show in Fig.\ref{Fig6} the numerical results obtained in this framework (see captions). The Gaussian input pulse becomes stimulated and the oscillating population is doubled every time it crosses the reservoir zone. The lifetime and LZT emission-induced losses become strongly compensated upon a relevant reservoir density as we can see in Fig.\ref{Fig6}(a). This figure should be compared to Fig.\ref{Fig6}(b) where the free oscillations (no reservoir is present) are displayed. We note that in (a) there is still a weak global decay of the number of particles, indeed our will is not to increase stimulate too much the density to avoid a switching to parametric instability\cite{FlayacBO,ReviewAtoms}. We have therefore created an almost persistent driven Bloch oscillations of polaritons despite of their lifetime as well as a maintained alternating spin emission of Fig.\ref{Fig3} with a single input pulse thanks to the bosonic and interacting nature of polaritons. This stimulation is not only advantageous from the point of view of the lifetime of the particles, but also because it induces a gain in the specific component that crosses the reservoir. Thus, it will tend to screen the deviations from a perfect $\sigma_\pm$ emission and further improves the efficiency of the device. We make the following final remark. A similar effect could also be achieved with a pure photonic system, moreover without strong disturbance from the excitonic reservoir. The amplification of the propagating wave would however require to achieve the gain condition, which occurs only with pumping powers typically one-two orders of magnitude larger than the amplification condition in a polaritonic system\cite{LasingThreshold,Ferrier2011}. From this point of view the use of the strong coupling is advantageous, whereas, on the other hand, it requires low temperature operation, at least in arsenide based systems.

\section{Conclusions}
We have analyzed spin dependent Bloch oscillations of exciton polaritons and photons. We have shown how the TE-TM splitting along the wire affects the spin dynamics of the particles during their motion. We have explained how the precession period of the pseudospin can be synchronized with the Bloch oscillations to reach a "Half Bloch-oscillations" regime. Capitalizing on this regime, a proposal for a spinoptronic emitter/converter and even a transistor based on the Landau-Zener tunneling has been made. Finally we have given two possible solutions to overcome the lifetime-induced signal losses.

\section{Acknowledgements.} We acknowledge the support of the joint CNRS-RFBR PICS project and of the FP7 ITN "Spin-Optronics" (Grant no. 237252).


\begin{thebibliography}{99}

\bibitem{Microcavities} A.V. Kavokin, J.J. Baumberg, G. Malpuech, F.P. Laussy, \emph{Microcavities}, Oxford University Press, (2007).

\bibitem{Savvidis} P.G. Savvidis, J. J. Baumberg, R. M. Stevenson, M. S. Skolnick, D. M. Whittaker, and J. S. Roberts, \emph{Phys. Rev. Lett.} \textbf{84}, 1547 (2000).

\bibitem{Stevenson} R.M. Stevenson, V. N. Astratov, M. S. Skolnick, D. M. Whittaker, M. Emam-Ismail, A. I. Tartakovskii, P. G. Savvidis, J. J. Baumberg, and J. S. Roberts, \emph{Phys. Rev. Lett.} \textbf{85}, 3680 (2000).

\bibitem{Baas} A. Baas, J.-Ph. Karr, M. Romanelli, A. Bramati, and E. Giacobino, \emph{Phys. Rev. B} \textbf{70}, 161307(R), (2004).

\bibitem{Gippiusb} N. A. Gippius , S. G. Tikhodeev , V. D. Kulakovskii , D. N. Krizhanovskii, and  A. I. Tartakovskii, \emph{Europhys. Lett.} \textbf{67}, 997 (2004).

\bibitem {Baumberg} J. J. Baumberg, P. G. Savvidis, R. M. Stevenson, A. I. Tartakovskii, M. S. Skolnick, D. M. Whittaker, and J. S. Roberts, \emph{Phys. Rev. B} \textbf{62}, R16247 (2000).

\bibitem{BECPolaritons}  J. Kasprzak, M. Richard, S. Kundermann, A. Baas, P. Jeambrun, J. M. J. Keeling, F. M. Marchetti, M. H. Szymanska, R. Andr\'{e}, J. L. Staehli, V. Savona, P. B. Littlewood, B. Deveaud, and Le Si Dang, \emph{Nature} \textbf{443}, 409 (2006).

\bibitem{PolaritonLaser} S. Christopoulos, G. Baldassarri H\"{ö}ger von H\"{ö}gersthal, A. J. D. Grundy, P. G. Lagoudakis, A. V. Kavokin, and J. J. Baumberg, \emph{Phys. Rev. Lett.} \textbf{98}, 126405 (2007).

\bibitem{AmoNatPhys}  A. Amo, J. Lefr\`{e}re, S. Pigeon, C. Adrados, C. Ciuti, I. Carusotto, R. Houdr\'{e}, E. Giacobino, and A. Bramati, \emph{Nat. Phy.} \textbf{5}, 805 (2009).

\bibitem{LagoudakisV} K. G. Lagoudakis, M. Wouters, M. Richard, A. Baas, I. Carusotto, R. Andr\'{e}, Le Si Dang, and B. Deveaud-Pl\'{e}dran, \emph{Nat. Phys.} \textbf{4}, 706 (2008).

\bibitem{AmoSoliton}  A. Amo, S. Pigeon, D. Sanvitto, V. G. Sala, R. Hivet, I. Carusotto, F. Pisanello, G. Lemenager, R. Houdre, E. Giacobino, C. Ciuti, and A. Bramati, \emph{arXiv}:1101.2530.

\bibitem{OSHEth}  A. Kavokin, G. Malpuech, and M. Glazov, \emph{Phys. Rev. Lett.} \textbf{95}, 136601 (2005).

\bibitem{OSHEexp}  C. Leyder, M. Romanelli, J. P. Karr, E. Giacobino, T. C. H. Liew, M. M. Glazov, A. V. Kavokin, G. Malpuech, and A. Bramati, \emph{Nat. Phys} \textbf{3}, 628 (2007).

\bibitem{Gippiusm} N. Gippius, I. Shelykh, D. Solnyshkov, A. Kavokin, Y. Rubo,, S. Gavrilov, S. Tikhoddev, G. Malpuech, \emph{Phys. Rev. Lett.} \textbf{98}, 236401 (2007).

\bibitem{Gippiusm2} D. Sarkar, S. S. Gavrilov, M. Sich, J. H. Quilter, R. A. Bradley, N. A. Gippius, K. Guda, V. D. Kulakovskii, M. S. Skolnick, and D. N. Krizhanovskii, \emph{Phys. Rev. Lett.} \textbf{105}, 216402 (2010).

\bibitem{Rubo} Y.G. Rubo, \emph{Phys. Rev. Lett.} \textbf{99}, 160401, (2007).

\bibitem{LagoudakisHV}  K. G. Lagoudakis, T. Ostatnick\'{y}, A. V. Kavokin, Y. G. Rubo, R. Andr\'{e} and B. Deveaud-Pl\'{e}dran, \emph{Science}, \textbf{326}, 974 (2009).

\bibitem{review_Shelykh} For a review on polaritons spin dynamics see: I. A. Shelykh, A. V. Kavokin, Y. G. Rubo, T. C. H. Liew, and G. Malpuech, \emph{Semicond. Sci. Technol.} \textbf{25} 013001 (2010).

\bibitem{Terahertz} K. V. Kavokin, M. A. Kaliteevski, R. A. Abram, A. V. Kavokin, S. Sharkova, and I. A. Shelykh, \emph{Appl. Phys. Lett.} \textbf{97}, 201111 (2010).

\bibitem{Neurons} T. C. H. Liew, A. V. Kavokin, and I. A. Shelykh, \emph{Phys. Rev. Lett.} \textbf{101}, 016402 (2008).

\bibitem{OpticalGate} C. Leyder, T. C. H. Liew, A. V. Kavokin, I. A. Shelykh, M. Romanelli, J. Ph. Karr, E. Giacobino, and A. Bramati, \emph{Phys. Rev. Lett.} \textbf{99}, 196402 (2007).

\bibitem{BerryPhase} I. A. Shelykh, G. Pavlovic, D. D. Solnyshkov, and G. Malpuech, \emph{Phys. Rev. Lett.} \textbf{102}, 046407 (2009).

\bibitem{DattaDas} I. A. Shelykh, R. Johne, D. D. Solnyshkov, and G. Malpuech, \emph{Phys Rev B} \textbf{82}, 153303 (2010).

\bibitem{Lausanne} T.K. Paraiso, M. Wouters, Y. Léger, F. Mourier-Genoud, and B. Deveaud-Pl\'{e}dran, \emph{Nature Materials} \textbf{9}, 655–660 (2010).

\bibitem{FlayacBO} H. Flayac, D. Solnyshkov, and G. Malpuech \emph{Phys. Rev. B} \textbf{83}, 045412 (2011).

\bibitem{TerahertzBOs} T. Dekorsy, P. Leisching, C. Waschke, K. Kohler, K. Leo, H. G. Roskos and H. Kurz, \emph{Semicond. Sci. Technol.} \textbf{9}, 1959 (1994).

\bibitem{CarusottoSForce} I. Carusotto, L. Pitaevskii, S. Stringari, G. Modugno, and M. Inguscio, \emph{Phys. Rev. Lett.} \textbf{95}, 093202 (2005).

\bibitem{Gravity} G. Ferrari, N. Poli, F. Sorrentino, and G. M. Tino, \emph{Phys. Rev. Lett.} \textbf{97}, 060402 (2006).

\bibitem{MalpuechBO} G. Malpuech, A. Kavokin, G. Panzarini, and A. Di Carlo, \emph{Phys. Rev. B} \textbf{63}, 035108 (2001).

\bibitem{AgarwalBO} V. Agarwal, J. A. del R\`{\i}o, G. Malpuech, M. Zamfirescu, A. Kavokin, D. Coquillat, D. Scalbert, M. Vladimirova, and B. Gil, \emph{Phys. Rev. Lett.} \textbf{92}, 097401 (2004).

\bibitem{Plagoud} M. Maragkou, C. E. Richards, T. Ostatnick\`{y}; J. D. Grundy Alastair; J. Zajac, M. Hugues, W. Langbein and P. G. Lagoudakis, \emph{Opt. Lett.} \textbf{36}, 1095 (2011).

\bibitem{MalpuechAPL} G. Malpuech, M. M. Glazov, I. A. Shelykh, P. Bigenwald, and K. V. Kavokin \emph{Appl. Phys. Lett.} \textbf{88}, 11118,(2006).


\bibitem{Maialle} M. Z. Maialle, E. A. de Andrada e Silva, and L. J. Sham \emph{Phys. Rev. B} \textbf{47}, 15776 (1993).

\bibitem{PolPatterns} W. Langbein, I. A. Shelykh, D. Solnyshkov, G. Malpuech, Yu. Rubo, and A. Kavokin, \emph{Phys. Rev. B} \textbf{75}, 075323 (2007).

\bibitem{FlayacHS} H. Flayac, D. D. Solnyshkov, and G. Malpuech, \emph{arXiv}:1103.4516, to appear in \emph{Phys. Rev. B}.

\bibitem{Klop} L. Klopotowski, M. D. Martin, A. Amo, L. Vina, I. A. Shelykh, M. M. Glazov, G. Malpuech, A.V. Kavokin and R. Andr\'{e}, \emph{Solid State Com.} \textbf{139}, 511 (2006).

\bibitem{Dasbach} G. Dasbach, A. A. Dremin, M. Bayer, V. D. Kulakovskii, N. A. Gippius, and A. Forchel, \emph{Phys. Rev. B} \textbf{65}, 245316 (2002).

\bibitem{Wertz} E. Wertz, L. Ferrier, D. D. Solnyshkov, R. Johne, D. Sanvitto, A. Lemaître, I. Sagnes, R. Grousson, A. V. Kavokin, P. Senellart, G. Malpuech and J. Bloch, \emph{Nature Physics} \textbf{6}, 860, (2010).

\bibitem{MagnussonJO} E. B. Magnusson, H. Flayac, I. A. Shelykh and G. Malpuech, \emph{Phys. Rev. B} \textbf{82}, 195312 (2010).

\bibitem{SavenkoDM} I. G. Savenko, E. B. Magnusson, and I. A. Shelykh, \emph{Phys. Rev. B} \textbf{83}, 165316 (2011).

\bibitem{ReviewAtoms} O. Morsh and M. Oberthaler, \emph{Rev. Mod. Phys.}, \textbf{78}, 179 (2006).

\bibitem{Keel} J. Keeling and N. G. Berloff, \emph{Phys. Rev. Lett.} \textbf{100}, 250401 (2008).

\bibitem{LasingThreshold} H. Deng, G. Weihs, D. Snoke, J. Bloch, and Y. Yamamoto, \emph{PNAS} \textbf{100} 26) 15318 (2003).

\bibitem{Ferrier2011} L. Ferrier, E. Wertz, R. Johne, D. D. Solnyshkov, P. Senellart, I. Sagnes, A. Lema\^{\i}tre, G. Malpuech, and J. Bloch, \emph{Phys. Rev. Lett.} \textbf{106} (2011).

\end{thebibliography}
\end{document}